\documentclass[preprint,longbibliography,superscriptaddress]{revtex4-1}
\usepackage{graphicx}
\usepackage{array}
\usepackage{amssymb}
\usepackage{amsfonts}
\usepackage{amsmath}
\usepackage{color}
\usepackage{booktabs}
\usepackage{threeparttable}
\usepackage{multirow}
\usepackage{subfigure}
\usepackage{epsfig}
\usepackage{threeparttable}
\usepackage{chngpage}
\usepackage{float}
\usepackage{color}
\usepackage{bm}
\usepackage{dcolumn}
\usepackage{textcomp}
\usepackage{verbatim}
\usepackage{epstopdf}
\usepackage{multirow}
\usepackage{tabularx}
\usepackage[labelfont=bf]{caption}
\DeclareCaptionLabelSeparator{vline}{ $|$ }
\usepackage{caption}
\makeatletter
\renewcommand\@biblabel[1]{#1.}
\makeatother

\begin{document}
\draft

\title{Quasiperiodicity and suppression of multistability in nonlinear dynamical systems}

\author{Ying-Cheng Lai} \email{Ying-Cheng.Lai@asu.edu}
\affiliation{School of Electrical, Computer and Energy Engineering, Arizona
State University, Tempe, Arizona 85287, USA}
\affiliation{Department of Physics, Arizona State University,
Tempe, Arizona 85287, USA}

\author{Celso Grebogi}
\affiliation{Institute for Complex Systems and Mathematical Biology,
King's College, University of Aberdeen, Aberdeen AB24 3UE, UK}

\date\today

\begin{abstract}

It has been known that noise can suppress multistability by dynamically 
connecting coexisting attractors in the system which are otherwise in
separate basins of attraction. The purpose of this mini-review is to 
argue that quasiperiodic driving can play a similar role in suppressing
multistability. A concrete physical example is provided where quasiperiodic 
driving was demonstrated to eliminate multistability completely to 
generate robust chaos in a semiconductor superlattice system. 

\end{abstract}

\maketitle

\section{Introduction} \label{sec:intro}

Professor Ulrike Feudel has made significant contributions, among many others, to two subfields
of nonlinear dynamics: quasiperiodicity~\cite{PF:1994,FPZ:1995,PF:1995,KPF:1995,FKP:1995,PZFK:1995,LFG:1996,FPP:1996,WFP:1997,FGO:1997,FWLG:1998,OF:2000,NSMF:2003,SPRF:2005,FKP:book}
and multistability~\cite{FGHY:1996,FG:1997,KFG:1999,KF:2002,KF:2003a,KF:2003b,FG:2003,NFS:2011,Pateletal:2014,PF:2014}. In particular, Ulrike was among the 
first~\cite{PF:1994,FPZ:1995,PF:1995,KPF:1995,FKP:1995,PZFK:1995}
to study strange nonchaotic attractors~\cite{DGO:1989}, attractors with a 
fractal geometry but without sensitive dependence on initial conditions, 
which arise commonly in quasiperiodically driven systems. While dynamical 
systems with multistability, i.e., systems with multiple coexisting 
attractors (each with a distinct basin of attraction), had been studied 
earlier in terms of the fractal structure of the basin boundaries and the 
physical consequences~\cite{GMOY:1983,MGOY:1985}, Ulrike was among the first 
to discover that even low-dimensional nonlinear dynamical 
systems can possess a large number of coexisting attractors~\cite{FGHY:1996}. 
The main purpose of this mini-review is to argue that there is a natural 
connection between the two subjects in the sense that quasiperiodicity
tends to suppress multistability. A physical example from the semiconductor 
superlattice system is quoted to support the argument.  

\begin{figure}
\centering
\includegraphics[width=\linewidth]{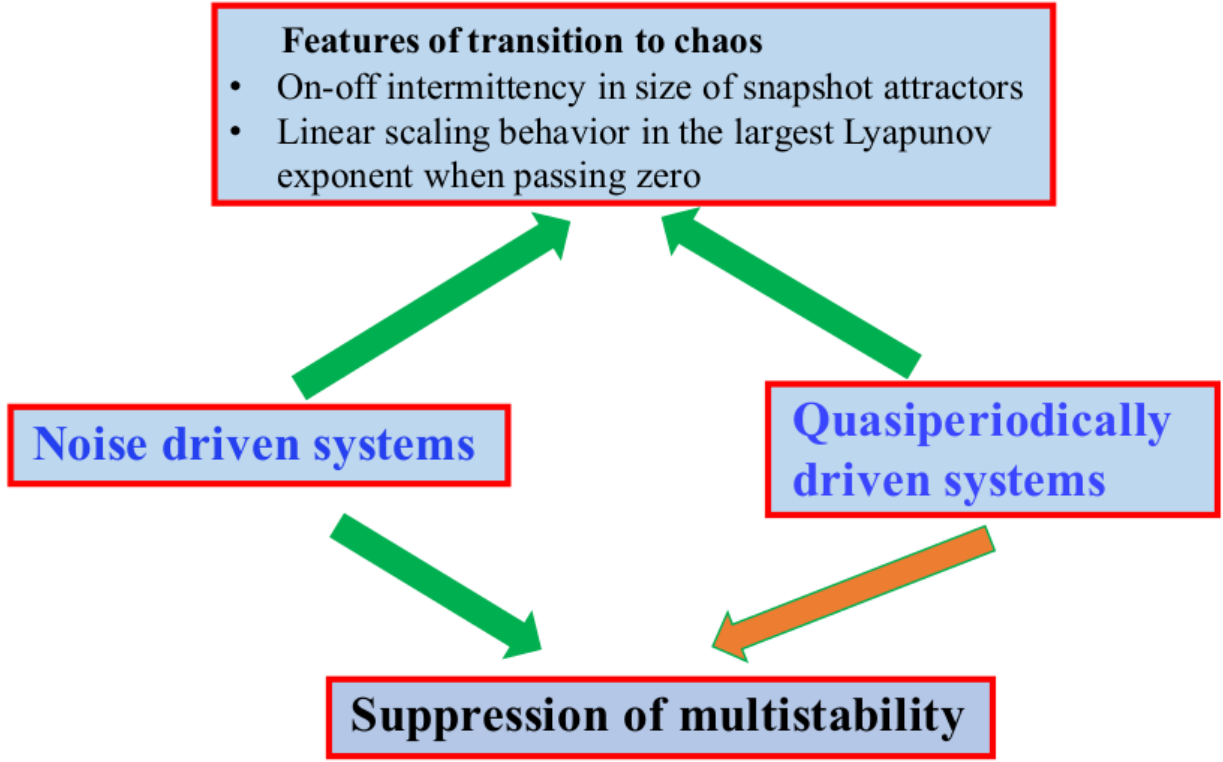}
\caption{ {\bf Reasoning that quasiperiodicity can suppress 
multistability.} Both random noise and quasiperiodic driving can induce
a transition to chaos with similar scaling behaviors. Since noise can 
suppress multistability, so should quasiperiodic driving. The green 
arrows indicate previously established scenarios. The lower-right arrow 
represents the proposition articulated in this mini-review.}
\label{fig:idea}       
\end{figure}

The line of reasoning leading to our proposition that quasiperiodicity 
can suppress or even diminish multistability is 
illustrated in Fig.~\ref{fig:idea}. It has  
been known that noise can induce transition to a chaotic attractor with 
distinct dynamical phenomena such as on-off intermittency in the size of
the snapshot attractors~\cite{YOC:1990,YOC:1991} formed by 
an ensemble of independent trajectories~\cite{RGO:1990}, and the linear 
scaling law of the largest Lyapunov exponent when it passes zero from the 
negative side~\cite{LLBS:2002,LLBS:2003,TL:2010}. The key observation 
underlying our argument is that these behaviors are shared by 
quasiperiodically driven systems about the transition to chaos~\cite{LFG:1996}.
In fact, in terms of transition to strange nonchaotic attractors, there is
also a striking resemblance in the scaling behaviors between situations
where the transition is due to quasiperiodic 
driving~\cite{Lai:1996,YL:1996,YL:1997} or noise~\cite{WZLL:2004}. 
Since noise can effectively suppress the number of attractors or stable states 
in the system by dynamically linking previously isolated 
attractors~\cite{FG:1997,Kaneko:1997,KFG:1999,KF:2002,FG:2003}, we expect 
that quasiperiodic driving can also be exploited to diminish 
multistability. Indeed, a recent study of the nonlinear dynamics of energetic 
electrons in a semiconductor superlattice system has revealed that 
multistability, which was present when the system is driven sinusoidally, 
can be effectively eliminated when an additional driving of non-commensurate 
frequency is introduced~\cite{YHL:2016}. 

\section{Transition to chaos} \label{sec:transition}

\subsection{Transition to a chaotic attractor in random dynamical systems}
\label{subsec:transition_rds}

A typical situation where noise-induced chaos can arise is periodic windows.
In such a window, there is a periodic attractor and a non-attracting chaotic 
set that leads to transient chaos~\cite{LT:book}. Noise can cause a trajectory
to visit both the original attractor and the non-attracting chaotic set,
giving rise to an extended noisy chaotic attractor. In a smooth dynamical 
system that exhibits chaos, in the absence of noise a chaotic attractor is
structurally unstable~\cite{LGYK:1993,KGPLS:2002,KG:2004}, whereby, the
periodic windows are dense and occupy
open sets in the parameter space. As a result, an arbitrarily small
perturbation can place the system in such a window, destroying the
chaotic attractor. When noise is present, chaos is enhanced in the
sense that noise-induced chaotic attractors can even occur in periodic 
windows. That is, chaotic attractors can now occur in open sets in the
parameter space. Noise-induced chaos thus provides the reason for
observing chaotic attractors in realistic systems. This phenomenon
is, of course, not restricted to periodic windows. Insofar as a nonchaotic 
attractor coexists with a non-attracting chaotic set in the phase space, a 
chaotic attractor can arise due to noise. In an opposite scenario, a 
chaotic attractor can be destroyed by noise through the mechanism of 
noise-induced crises~\cite{SOG:1991}. There are also situations where 
a chaotic attractor, which is typically nonhyperbolic, is robust
against noise~\cite{KGPLS:2002}.

A standard approach to defining a chaotic attractor under noise  
is the sensitive dependence on initial conditions, as characterized by 
the existence of at least one positive Lyapunov 
exponent~\cite{LY:1988,YOC:1990}. This is because the Lyapunov exponents 
are the time-averaged stretching or contracting rates of infinitesimal 
vectors along a typical trajectory in the phase space, which can be defined 
for both deterministic and stochastic dynamical systems. In particular, in 
the absence of noise, if the attractor is not chaotic, the largest 
Lyapunov exponent of the asymptotic attractor is negative for 
maps (zero for flows). As noise is turned on and its strength becomes 
sufficiently large, there is a nonzero probability that a trajectory 
originally on the attracting set escapes it and wanders near the coexisting
non-attracting chaotic set. In this case, the largest Lyapunov exponent 
$\lambda_1$ becomes positive, indicating that the asymptotic attractor of 
the system is chaotic for trajectories starting from random initial 
conditions.

A basic issue concerns the scaling of the largest Lyapunov exponent about 
the transition~\cite{LLBS:2002,LLBS:2003,TL:2010}. Consider, for 
example, a continuous-time dynamical system. When the system
is in a periodic window and exhibits a periodic attractor, in the
deterministic case the largest Lyapunov exponent $\lambda_1$ is zero.
As the noise amplitude $\sigma$ is increased from zero and passes
through a critical point $\sigma_c$, $\lambda_1$ becomes positive.
It has been argued and supported by numerical simulations~\cite{LLBS:2002}
that $\lambda_1$ obeys the following scaling law with the variation
in the noise amplitude beyond the critical value:
\begin{equation} \label{eq:scaling}
\lambda_1 \sim (\sigma - \sigma_c)^{\alpha},
\end{equation}
for $\sigma > \sigma_c$, where $\alpha = 1$. There were heuristic 
arguments~\cite{LLBS:2002,LLBS:2003} leading to (\ref{eq:scaling}),
which were based on analyzing the overlaps between the natural measure 
of the noise-enlarged periodic attractor and that of the stable manifold 
of the non-attracting chaotic set. A theoretical approach to deriving
(\ref{eq:scaling}) based on the concept of {\em quasipotentials}~\cite{KF:2003a,KF:2003b,FW:book,GT:1984,Grah:1989,Gras:1989,GHT:1991,RT:1991,HG:1992,HTG:1994,KG:2004} was also developed~\cite{TL:2010}.

In terms of geometry, without noise, a chaotic attractor
typically exhibits a fractal structure caused by the underlying dynamics.
Under the influence of small random perturbations, if one examines a
long trajectory produced by the dynamics, one usually observes that the
fractal structure is smeared up to a distance scale proportional to
the strength of the perturbations. In order to observe a clear fractal
structure, a remedy is to examine the snapshot pattern formed by an ensemble 
of trajectories subject to the {\em same} random perturbation~\cite{RGO:1990}.
The details of the fractal structure differ from time to time, but
the fractal dimensions {\em remain invariant}~\cite{LY:1988,RGO:1990}.
The idea of {\em snapshot attractors}~\cite{RGO:1990} was useful 
for visualizing and characterizing fractal patterns arising in physical 
situations such as passive particles advected on the surface of 
a fluid~\cite{YOC:1991,SO:1993}. It was found~\cite{YOC:1990,YOC:1991} that 
slightly after the transition to chaos, the size of the snapshot attractor 
can exhibit an extreme type of intermittency - on-off 
intermittency~\cite{FY:1985,FY:1986,FIIY:1986,PST:1993,HPH:1994,PHH:1994,HPHHL:1994,SO:1994,LG:1995,VAOS:1995,AS:1996,Lai:1996a,Lai:1996b,VAOS:1996,MALK:2001,RC:2007,HNDSCL:2017}.
Snapshot attractors can also be used to study nonstationary dynamical
systems~\cite{SLC:2008,KFT:2016}.

\subsection{Transition to a chaotic attractor in quasiperiodically driven
systems} \label{subsec:transition_qpfds}

In a quasiperiodically driven dynamical system, it was 
demonstrated~\cite{LFG:1996} that, similar to the transition to chaos in random
dynamical systems, the largest nontrivial Lyapunov exponent passes through 
zero {\it linearly} with the parameter about the transition. In fact,
near the transition, the tangent vector along a typical trajectory 
experiences both time intervals of expansion and time intervals of contraction.
On the nonchaotic side, the Lyapunov exponent is slightly negative and,
hence, contraction dominates over expansion. On the chaotic side where the 
Lyapunov exponent is slightly positive, expansion dominates over contraction.
A consequence is that the collective behavior of
an ensemble of trajectories observed at different instants of times
exhibits an extreme type of intermittency on the chaotic side of
the transition. During the expansion time intervals, the trajectories
burst out by separating from each other, but during the contraction
time intervals the trajectories merge together. Therefore, if one examines 
the snapshot attractors of this ensemble of trajectories at different
times, one finds that the size of the chaotic attractor varies drastically in 
time in an intermittent fashion. The average size of the snapshot attractor
scales {\it linearly} with a parameter above but near the transition.
In addition, the average interval between bursts also scales linearly 
with the parameter above the transition.

In Ref.~\cite{LFG:1996}, the following quasiperiodically forced damped 
pendulum system~\cite{RBOAG:1987} was considered:
\begin{equation} \label{eq:qfp_1}
\frac{d^2\theta}{dt^2} + \nu\frac{d\theta}{dt} + \sin{\theta}
= K + V[\cos{(\omega_1 t)} + \cos{(\omega_2 t)}], 
\end{equation}
where $\theta$ is the angle of the pendulum with the vertical axis,
$\nu$ is the dissipation rate, $K$ is a constant, $V$ is the forcing 
amplitude, $\omega_1$ and $\omega_2$ are the two incommensurate frequencies. 
Introducing two new variables, $t \rightarrow \nu t'$ and 
$\phi \equiv \theta + \pi/2$, Eq.~(\ref{eq:qfp_1}) becomes (after 
setting $t' = t$)
\begin{displaymath}
\frac{1}{p}\frac{d^2\phi}{dt^2} + \frac{d\phi}{dt} - \cos{\phi} 
= K + V[\cos{(\omega_1 t)} + \cos{(\omega_2 t)}],
\end{displaymath}
where $p = \nu^{2}$ is a new parameter, $\omega_1$ and $\omega_2$
have been rescaled accordingly: $\omega_1\rightarrow \omega_1\nu$
and $\omega_2\rightarrow \omega_2\nu$. In terms of the dynamical variables 
$\phi$, $v\equiv d\phi/dt$, and $z\equiv \omega_2 t$, one has
\begin{eqnarray}
 \frac{d\phi}{dt} & = & v, \nonumber \\
 \frac{dv}{dt}    & = & p [K + V(\cos{(\frac{\omega_1}{\omega_2}z)} + \cos{z})
+ \cos{\phi} - v], \nonumber \\
 \frac{dz}{dt}    & = & \omega_2. \label{eq:autonomous}
\end{eqnarray}
The quasiperiodically driven system Eq.~(\ref{eq:autonomous}) exhibits rich 
dynamical phenomena~\cite{RBOAG:1987,RO:1987}. For example, for different 
parameters in the $K-V$ plane, one finds two- and three-frequency quasiperiodic
attractors, strange nonchaotic attractors and chaotic attractors emerging 
from two-frequency quasiperiodic attractors. In the strong damping limit 
$p\rightarrow\infty$, Eq.~(\ref{eq:autonomous}) reduces to a first-order
equation which is isomorphic to the Schr\"{o}dinger equation with
a quasiperiodic potential~\cite{BOA:1985}.

\begin{figure}
\centering
\includegraphics[width=0.8\linewidth]{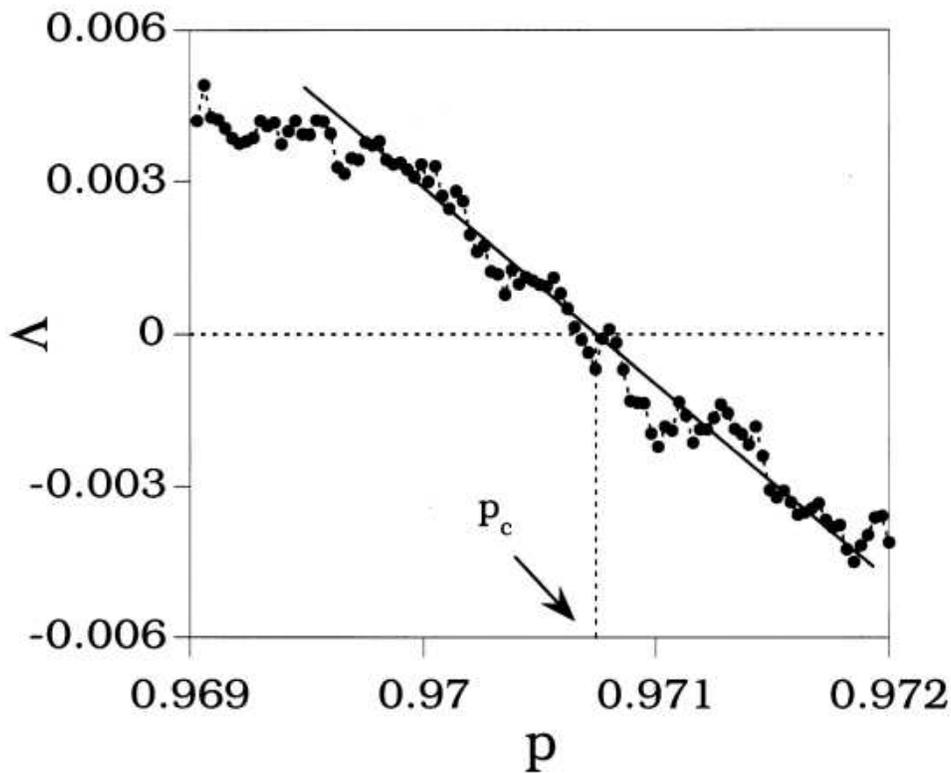}
\caption{ {\bf Linear behavior of Lyapunov exponent about the transition
to chaos in the quasiperiodically driven pendulum system}. For system 
described by Eq.~(\ref{eq:autonomous}), the largest nontrivial Lyapunov 
exponent $\Lambda$ versus the bifurcation parameter $p$ (damping rate) 
for $ 0.969 \le p \le 0.972$. Other parameter values are $V = 0.55$, 
$K = 0.8$, $\omega_2 = 1$ and $\omega_1$ is the inverse golden mean.
{\em From Ref.~\cite{LFG:1996} with permission}.}
\label{fig:Lya}
\end{figure}

\begin{figure}
\centering
\includegraphics[width=0.68\linewidth]{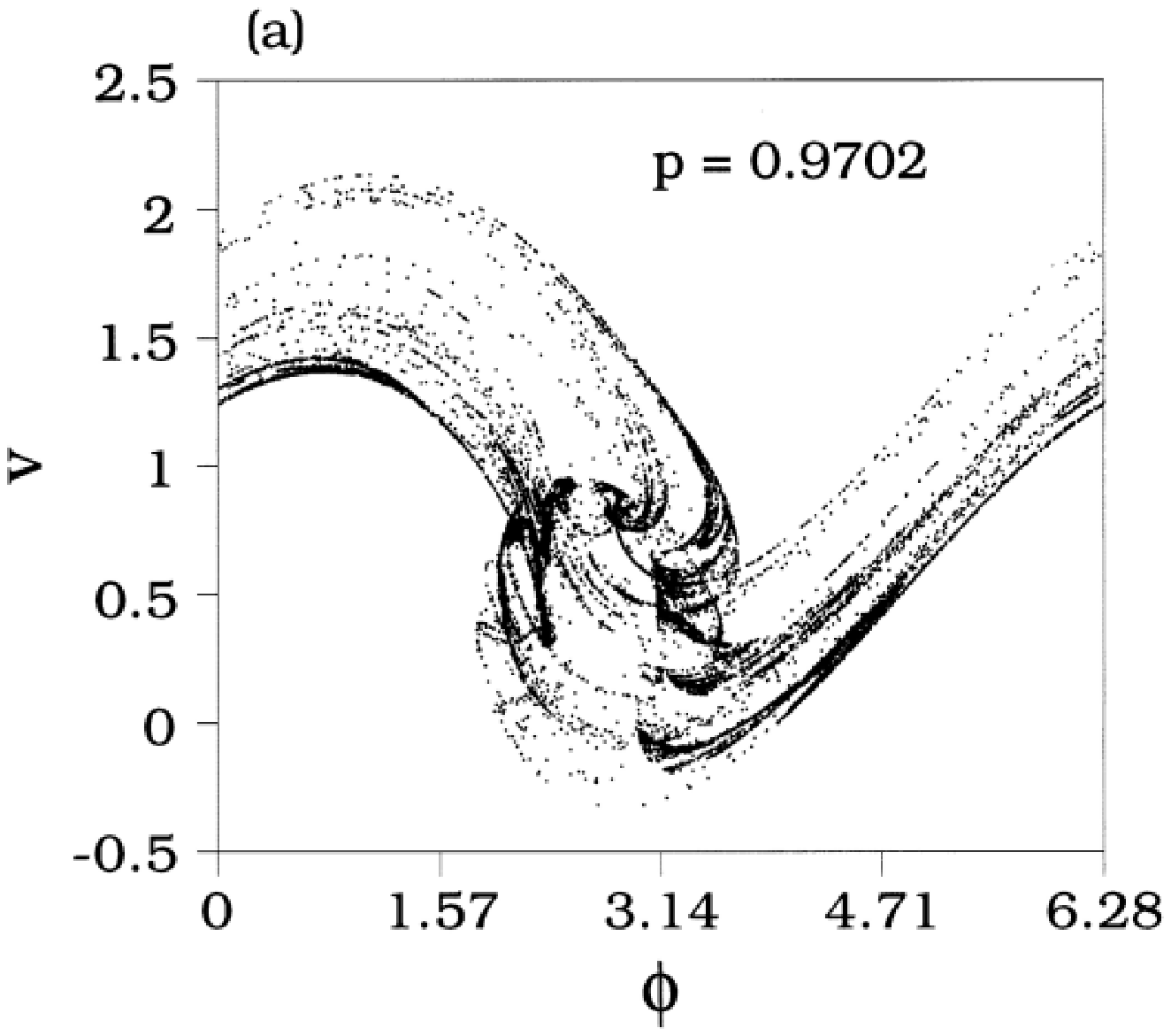}
\includegraphics[width=0.68\linewidth]{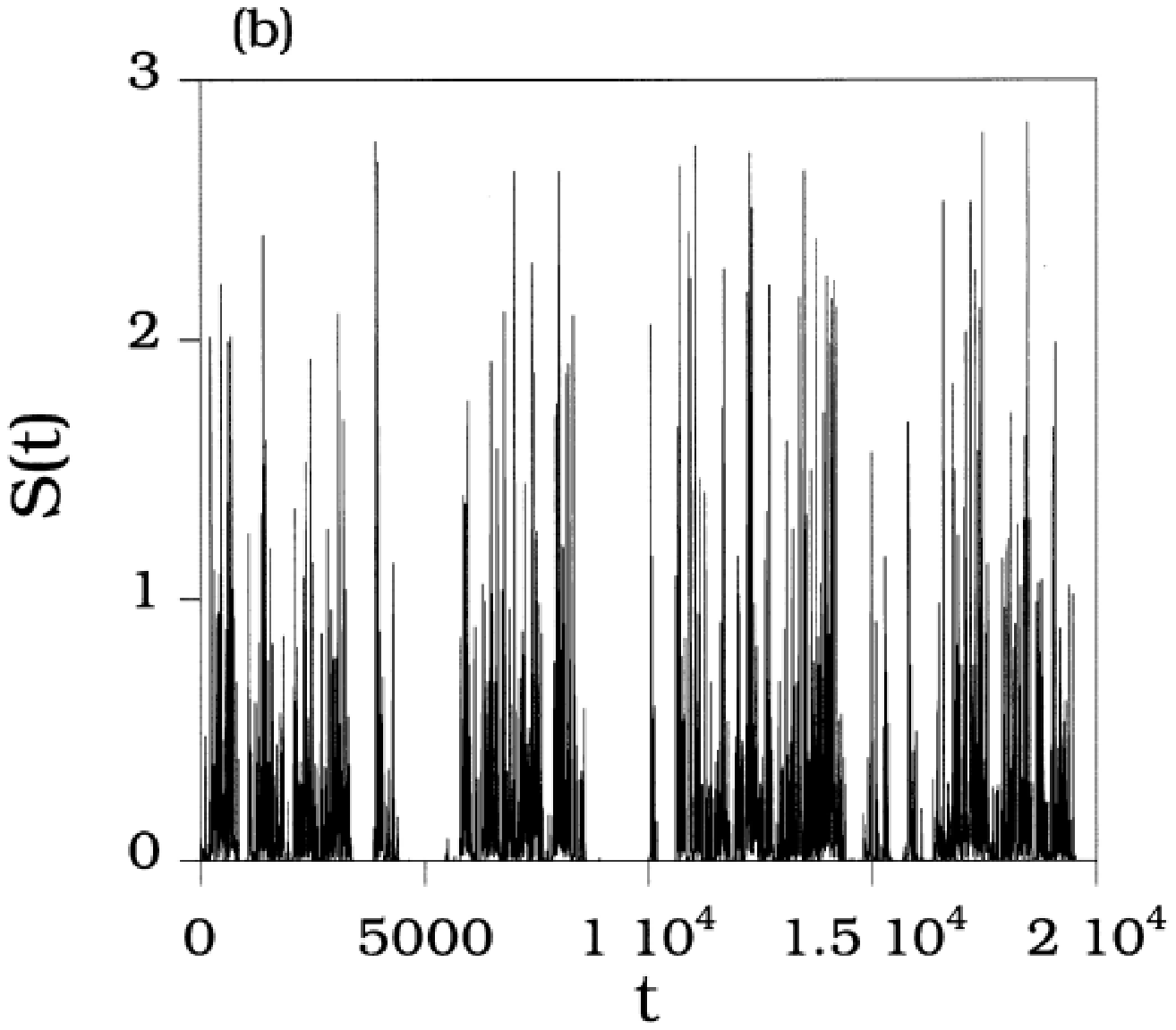}
\caption{ {\bf Behavior of the snapshot attractor about the transition
to chaos in the quasiperiodically driven pendulum system}. For system 
described by Eq.~(\ref{eq:autonomous}), (a) a single trajectory on the 
chaotic attractor for $p = 0.9702$ ($\Lambda\approx 0.002$). (b) $S(t)$, 
the size of the snapshot attractors computed using 128 trajectories, 
versus time $t$ for $p = 0.9702$. 
{\em From Ref.~\cite{LFG:1996} with permission}.}
\label{fig:snapshot}
\end{figure}

In Ref.~\cite{LFG:1996}, numerical results were presented for $K = 0.8$, 
$V = 0.55$, $\omega_1 = (\sqrt{5} - 1)/2$ (the inverse golden mean), 
$\omega_2 = 1.0$, with $p$ as the bifurcation parameter. For large 
values of $p$ ($p>1.0$), the damping is strong so that the motion is typically
periodic or quasiperiodic. As $p$ is decreased, say, $p < p_{c} \approx 1.0$, 
both strange nonchaotic and chaotic attractors exist. Figure~\ref{fig:Lya} 
shows the largest nontrivial Lyapunov exponent $\Lambda$ for 
$p\in [0.969, 0.972]$. The transition occurs at $p = p_c \approx 0.9707$ 
where $\Lambda >0$ for $p<p_c$ and $\Lambda <0$ for $p>p_c$. The variables 
$\phi$ and $v$ on the stroboscopic surface of section defined by 
$z = n(2\pi), n = 0, 1, \ldots$ 
can be used to visualize the attractors. Figure~\ref{fig:snapshot}(a) 
shows a single long trajectory on the chaotic attractor for $p = 0.9702 < p_c$
($\Lambda\approx 0.002$). Examination of the attractors for $p$ slightly above
$p_c$ indicates that they are strange nonchaotic~\cite{LFG:1996}. One feature 
about the transition is that the Lyapunov exponent $\Lambda$ passes through 
zero linearly, in spite of fluctuations caused by finite length of trajectories 
used in numerical computation. The mechanism behind such a smooth transition 
can be understood by examining the relative weights of the phase-space 
regions where a typical trajectory experiences expansion and 
contraction~\cite{Lai:1996}.

The time evolution of the snapshot attractor was also studied~\cite{LFG:1996}.
Specifically, an ensemble of initial conditions on the $[\phi,v]$ plane was
chosen with the same $z(0) = 0$ (so they start to evolve at the same time). 
Snapshot attractors, $i.e.$, the distribution of the trajectories resulting 
from these initial conditions in the phase space at fixed subsequent 
instants of time, were examined. It was found that the properties of the 
snapshot attractors are qualitatively different for
$p$ slightly above $p_c$ ($\Lambda$ slightly negative) 
and $p$ slightly below $p_c$ ($\Lambda$ slightly positive).
In particular, for $p$ slightly above $p_{c}$ on the nonchaotic side,
the trajectories resulting from these initial conditions eventually
converge to a single trajectory. At any instant of time (after sufficiently
long transient time), the snapshot attractor of these trajectories consists of
only one point in the phase space. As time progresses, the single point
for all trajectories moves in the phase space, tracing out a trajectory
which lies on the strange nonchaotic attractor. The time required for the 
ensemble of trajectories to converge to a single trajectory scales as 
$\tau \sim 1/|\Lambda| \sim 1/|p - p_c|$. However, an intermittency behavior 
occurs on the chaotic side for $p$ slightly below $p_c$ with $\Lambda$ being 
slightly positive. In this case, the snapshot attractors are no longer a 
single point even after a long transient time. There are time intervals 
during which the snapshot attractor consists of points spread over the 
entire chaotic attractor. There are also time intervals during which the 
snapshot attractor consists of points concentrated on extremely 
small regions in the phase space. The {\it size} of the snapshot attractors, 
therefore, changes drastically with time in an intermittent fashion.

To quantify intermittency, the following time-dependent size of the 
snapshot attractors can be examined:
\begin{equation}
S(t) = \sqrt{\frac{1}{N}\sum_{i=1}^{N}\{[\phi_{i}(t) - \langle\phi(t)\rangle]^2 
+ [v_i(t) - \langle v(t)\rangle]^2\}}, \label{eq:size}
\end{equation}
where $N$ is the number of points on the snapshot attractors,
[$\langle\phi(t)\rangle$, $\langle v(t)\rangle$] 
defines the geometric center of these points at a given time: 
$\langle \phi(t)\rangle \equiv \frac{1}{N}\sum^{N}_{i=1}\phi_{i}(t)$ and
$\langle v(t)\rangle \equiv \frac{1}{N}\sum^{N}_{i=1}v_i(t)$.
Figure~\ref{fig:snapshot}(b) shows $S(t)$ versus $t$ for $p = 0.9702$,
where $t$ is the integer time measured on the surface
of section corresponding to the real time $t(2\pi/\omega_2)$,
and the snapshot attractors are computed from 128 initial conditions
uniformly chosen along the diagonal line of the rectangle defined by
$\phi(0)\in[0,2\pi]$ and $v(0)\in [-1,1]$. It can be seen that the
size of the snapshot attractors exhibits an extreme type of intermittent
behavior, the so-called ``on-off'' intermittency~\cite{FY:1985,FY:1986,FIIY:1986,PST:1993,HPH:1994,PHH:1994,HPHHL:1994,SO:1994,LG:1995,VAOS:1995,AS:1996,Lai:1996a,Lai:1996b,VAOS:1996,MALK:2001,RC:2007,HNDSCL:2017}.
There are time intervals during which the snapshot attractors are concentrated
on a region with extremely small size ($< 10^{-14}$). The time averaged size 
of the snapshot attractors {\it on the chaotic side} near the transition, 
defined as
$\langle S(t)\rangle = \lim_{T\rightarrow\infty} \frac{1}{T}\int_{0}^{T}S(t)dt$,
obeys the following scaling relation~\cite{LFG:1996}:
\begin{equation}
\langle S(t)\rangle \sim \Lambda \sim |p - p_c|.
\end{equation}
In fact, for $p$ slightly above $p_c$ on the nonchaotic side, the averaged 
size of the snapshot attractor is quite small, yet nonzero. This is due 
to the finite simulation time. As the computation time increases, the 
averaged size decreases towards zero. 

\section{Destruction of multistability by quasiperiodic driving in 
a semiconductor superlattice system} \label{sec:superlattice}

We demonstrated in Sec.~\ref{sec:transition} some common features and 
scaling behaviors associated with transition to chaos in noise and 
quasiperiodically driven systems. The implication is that, since noise 
can suppress multistability, quasiperiodic driving should have a similar
effect. To support this idea, we present a concrete physical system - a 
semiconductor superlattice, to demonstrate that quasiperiodic driving
can eliminate multistability and lead to robust chaos.    

A semiconductor superlattice consists of a periodic sequence of thin
layers of different types of semiconductor materials~\cite{ET:1970} 
so that the electronic properties of the structure can be engineered. 
Specifically, a superlattice is a periodic structure of coupled quantum wells,
where at least two types of semiconductor materials with different band gaps
are stacked on top of each other along the so-called growth direction
in an alternating fashion~\cite{Grahn:book,ZKKPG:1996}. 
An effective approach to modeling transport dynamics in the superlattice
system is through the force-balance equation~\cite{LT:1984,LT:1985,LHC:1991,Lei:1994a,Lei:1994b,IDS:1991,Gerhardts:1993,BT:1977,BT1:1979,BT2:1979},
which can be derived either from the classical Boltzmann transport
equation~\cite{Lei:1994a,Lei:1994b} or from the Heisenberg equation of
motion~\cite{Lei:1992,Lei:1995}. In spite of a quantum system's being
fundamentally linear, the self-consistent field caused by the combined
effects of the external bias and the intrinsic many-body mean field
becomes effectively nonlinear~\cite{ACMKC:1998,ABCCC:1996}. In the high field
transport regime, various nonlinear phenomena including chaos can
arise~\cite{Scholl:book}. In the past two decades, there were a host
of theoretical and computational studies of chaotic dynamics in
semiconductor superlattices~\cite{Scholl:book,ABCCC:1996,BGB:1996,PSS:1998,ACMKC:1998,LGPB:1998,ACMKC2:1998,CL:1999,CLL:2000,SPB:2001,AK:2002,ABC:2002,ASWS:2002,BG:2005,HS:2008,GBSF:2009}. The effects of magnetic field on the
nonlinear dynamics in superlattices were also
investigated~\cite{Stapletonetal:2004,WC:2005,BFPEF:2008}. Experimentally,
a number of nonlinear dynamical behaviors were observed and
characterized~\cite{LGTP:1998,ZKKPG:1996,ZKGP:1997,Fromholdetal:2004,Hramovetal:2014}.

A key application of semiconductor superlattices is to fill the so-called
``THz'' gap, i.e., to develop radiation sources, amplifiers and
detectors~\cite{HSA:2007,HAT:2008,HAMA:2009,HMA:2009,Hyart:book}
from 0.1 to 10 THz, the frequency range in which convenient
radiation sources are not readily available~\cite{Siegel:2002,FZ:2002,CBPHW:2005,Tonouchi:2009}. In particular,
below 0.1 THz electron transport based devices are typical, and above
10 THz devices based on optical transitions (e.g., solid state lasers)
are commonly available. Since in general, chaotic systems can be used
as random number generators~\cite{Kocarev:2001,SK:2001,SPK:2001,DG:2006,LC:2006,Uchidaetal:2008,RARK:2009,CHLGD:2010,ZWLXLWZ:2012},
ubiquity of chaos in semiconductor superlattices implies that such systems
may be exploited for random signal generation in the frequency range
corresponding to the THz gap. 

In a recent paper~\cite{YHL:2016}, the 
dynamics of energetic or ``hot'' electrons in semiconductor superlattices 
were investigated. Specifically, the setting was studied where the 
system is subject to strong $dc$ and $ac$ fields so that
dynamical resonant tunneling occurs effectively in a quasi-one-dimensional
superlattice. Due to the strong driving field, a space charge field
is induced, which contains two nonlinear terms in the equation of motion. 
In particular, using the force-balance equation~\cite{HAAC:2005} for an 
$n$-doped semiconductor quantum-dot superlattice, the dynamical equation
for the electron center-of-mass velocity $V_{\rm c}(t)$ can be written as
\begin{equation} \label{eq:COM_v}
\frac{dV_{\rm c}(t)}{dt} = -\left[\gamma_1+\Gamma_{\rm c}
\sin{(\Omega_{\rm c}t)}\right]\,V_{\rm c}(t) 
+ \frac{e}{M({\cal E}_{\rm e})}\left[E_0+E_1
\cos{(\Omega_1t)} + E_1^\prime\cos{(\Omega^\prime_1t)} + E_{\rm sc}(t)\right],
\end{equation}
where $\gamma_1$ is the momentum-relaxation rate constant, $\Gamma_{\rm c}$
comes from the channel-conductance modulation with $\Omega_{\rm c}$ being
the modulation frequency, $M({\cal E}_{\rm e})$ is the energy-dependent
averaged effective mass of an electron in the superlattice,
${\cal E}_{\rm e}(t)$ is the average energy per electron, $E_0$ is the applied
$dc$ electric field, $E_1$ and $E_1^\prime$ are the amplitudes of the two
external $ac$ fields with frequencies $\Omega_1$ and $\Omega_1^\prime$,
respectively, and $E_{\rm sc}(t)$ is the induced space-charge field due
to the excitation of plasma oscillation. Here, the statistical resistive
force~\cite{HAAC:2005} has been approximated by the momentum relaxation rate.
Based on the energy-balance equation, one can show~\cite{HA:2008} that
${\cal E}_{\rm e}(t)$ satisfies the following dynamical equation
\begin{equation} \label{eq:E_e}
\frac{d{\cal E}_{\rm e}(t)}{dt}=-\gamma_2\left[{\cal E}_{\rm e}(t)
-{\cal E}_0\right] 
+eV_{\rm c}(t)
\left[E_0+E_1\cos(\Omega_1t)+E_1^\prime\cos(\Omega^\prime_1t)
+E_{\rm sc}(t)\right],
\end{equation}
where $\gamma_2$ is the energy-relaxation rate constant and ${\cal E}_0$
is the average electron energy at the thermal equilibrium, and
the thermal energy exchange of the electrons with the crystal
lattice~\cite{HA:2008} is approximately described by the $\gamma_2$ term.
Applying the Kirchoff's theorem to a resistively shunted quantum-dot
superlattice~\cite{ACMKC:1998}, one obtains~\cite{HC:2003} the dynamical
equation for the induced space-charge field $E_{\rm sc}(t)$ as
\begin{equation} \label{eq:SC_field}
{\frac{dE_{\rm sc}(t)}{dt}=-\gamma_3\,E_{\rm sc}(t)
-\left(\frac{en_0}{\epsilon_0\epsilon_b}\right)V_{\rm c}(t)}\ ,
\end{equation}
where $\gamma_3$, which is inversely proportional to the product of the
system resistance and the quantum capacitance, is the dielectric relaxation
rate constant~\cite{HC:2003}, $n_0$ is the electron concentration at
the thermal equilibrium, and $\epsilon_b$ is the relative dielectric
constant of the host semiconductor material. The exact microscopic
calculations of $\gamma_1$ and $\gamma_2$ in the absence of space-charge field
were carried out previously~\cite{HLG:2009} based on the semiclassical
Boltzmann transport equation and the coupled force-energy balance
equations~\cite{CLL:2000}, respectively. Equivalent quantum calculations of
$\gamma_1$ and $\gamma_2$ can also be done through the coupled force
balance and the Boltzmann scattering equations~\cite{HAAC:2005}.

Within the tight-binding model, the single-electron kinetic energy
$\varepsilon_k$ in a semiconductor quantum-dot superlattice can be written as
\begin{equation} \label{eq:KE}
\varepsilon_k=(\Delta/2)\left[1-\cos(kd)\right],
\end{equation}
where $k$ ($|k|\leq \pi/d$) is the electron wave number along the superlattice
growth direction, $\Delta$ is the miniband width, and $d$ is the spatial period
of the superlattice. This energy dispersion relation gives~\cite{HAAC:2005}
\begin{equation} \label{eq:M_E}
\frac{1}{M({\cal E}_{\rm e})}=\left\langle\frac{1}{\hbar^2}\,
\frac{d^2\varepsilon_k}{dk^2}\right\rangle=\frac{1}{m^\ast}
\left[1-\left(\frac{2}{\Delta}\right){\cal E}_{\rm e}(t)\right]\ ,
\end{equation}
where $m^\ast=2\hbar^2/\Delta d^2$ and $|1/M({\cal E}_{\rm e})|\leq 1/m^\ast$.

\begin{figure}
\centering
\includegraphics[width=0.8\linewidth]{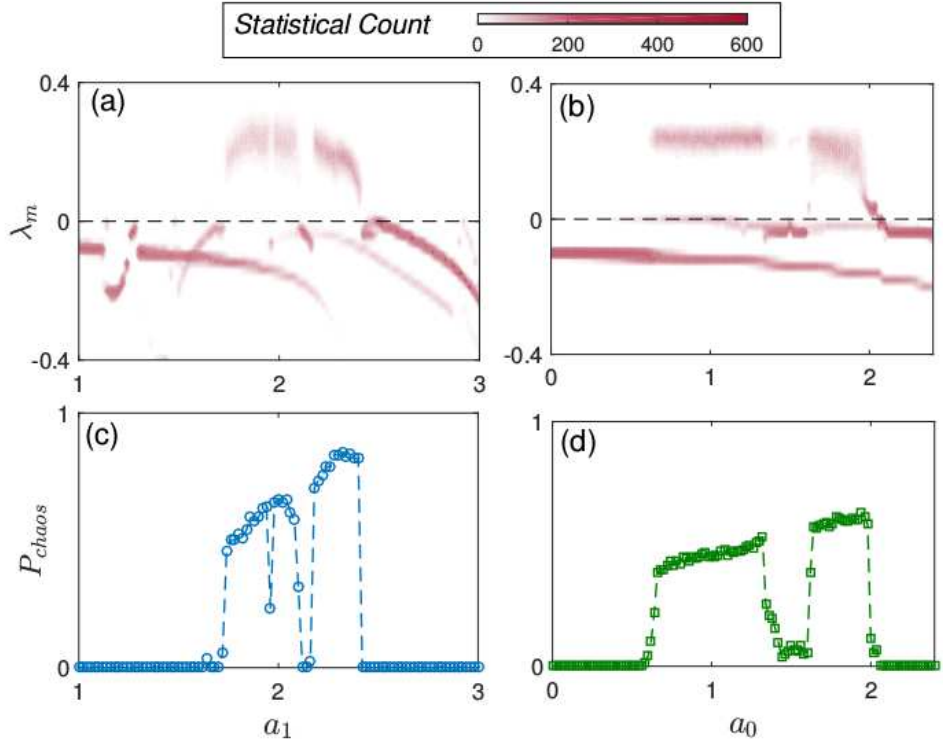}
\caption{ {\bf Transition to chaos and multistability}.
(a) For fixed $a_0 = 2.23$, the values of the maximum Lyapunov exponent
$\lambda_m$ calculated from an ensemble of initial conditions versus
$a_1$ for $1.0 \leq a_1 \leq 3.0$ . (b) A similar plot but for fixed
$a_1 = 2.13$ and $a_0$ varying in the range $[1.0,2.4]$. (c) For $a_0 = 2.23$,
the probability versus $a_1$ for a random trajectory to land in a chaotic
attractor. (d) A plot similar to that in (c) but for fixed $a_1 = 2.13$ and
varying $a_0$. Other parameters are $a^\prime_1=a_2=0$, $a_3=7.48$,
$b_1=0.28$, $b_2=b_3=2.85\times 10^{-2}$, and $\Omega=1.34$.
From (a) and (c), abrupt emergence
of chaos at $a_1 \approx 1.65$ and abrupt disappearance of chaos at
$a_1 \approx 2.45$ can be seen (see text for the reason of the ``abruptness'').
The dips in the probability curve of chaos for $a_1 \approx 2.0$ and
$a_1 \approx 2.15$ are due to periodic windows. Abrupt emergence and
disappearance of multistability associated with chaos also occur for
fixed $a_1 = 2.13$ and varying $a_0$, as shown in (b) and (d). 
{\em From Ref.~\cite{YHL:2016} with permission}.}
\label{fig:SL_p}
\end{figure}

For numerical calculations, it is convenient to use dimensionless quantities:
$v(\tau)=(m^\ast d/\hbar)\,V_{\rm c}$,
$w(\tau)=[(2/\Delta)\,{\cal E}_{\rm e}-1]$,
$f(\tau)=(ed/\hbar\omega_0)\,E_{\rm sc}$, and $\tau=\omega_0t$ with
$\omega_0=1$\,THz being the frequency scale. In terms of the dimensionless
quantities, the dynamical equations of the resonantly tunneling electrons in
the superlattice become
\begin{eqnarray} \label{eq:dlesseq}
\frac{dv(\tau)}{d\tau} & = & -b_1v(\tau)\left[1+a_2\sin(\bar{\Omega}\tau)\right]\\ \nonumber
&-&\left[a_0+a_1\cos(\Omega\tau)
+a^\prime_1\cos(\Omega^\prime\tau)+f(\tau)\right]w(\tau), \\ \nonumber 
\frac{dw(\tau)}{d\tau} &=& -b_2[w(\tau)-\overline{w}_0] 
+\left[a_0+a_1\cos(\Omega\tau)
+a^\prime_1\cos(\Omega^\prime\tau)+f(\tau)\right]v(\tau), \\ \nonumber
\frac{df(\tau)}{d\tau} &=& -b_3f(\tau)-a_3v(\tau),
\end{eqnarray}
where $\overline{w}_0=[(2/\Delta)\,{\cal E}_0-1]=-1$, $b_1=\gamma_1/\omega_0$,
$b_2=\gamma_2/\omega_0$, $b_3=\gamma_3/\omega_0$, $a_0=\omega_B/\omega_0$,
$a_1=\omega_s/\omega_0$, $a^\prime_1=\omega^\prime_s/\omega_0$,
$a_2=\Gamma_{\rm c}/\gamma_1$ and $a_3=(\Omega_c/\omega_0)^2$ are all
positive real constants. The field related parameters are
$\omega_B=eE_0d/\hbar$, $\omega_s=eE_1d/\hbar$,
$\omega^\prime_s=eE^\prime_1d/\hbar$,
$\Omega=\Omega_1/\omega_0$, $\Omega^\prime=\Omega^\prime_1/\omega_0$,
$\bar{\Omega}=\Omega_{\rm c}/\omega_0$, and
$\Omega_c=\sqrt{e^2n_0/m^\ast\epsilon_0\epsilon_b}$, where the last quantity
is the bulk plasma frequency. The fields are assumed to be turned on at $t=0$.
The initial conditions for Eq.~(\ref{eq:dlesseq}) are $v(0)=v_0$, $f(0)=f_0$
and $w(0)=w_0$.

The issue addressed~\cite{YHL:2016} was that of reliability and robustness
of chaos, i.e., for a given parameter setting, what is the probability to 
generate chaos from a random initial condition? It was found that, for the 
common case of a single $ac$ driving field, onset of chaos is typically 
accompanied by the emergence of multistability in the sense that there are 
coexisting attractors in the phase space which are not chaotic. Using the 
ensemble method to calculate the maximum Lyapunov exponent, the motions on the
regular and chaotic attractors can be distinguished. The probability for
a random initial condition to lead to chaos is finite but in general
is not close to unity. Due to the simultaneous creation of the basin
of attraction of the chaotic attractor, the transition to multistability
with chaos, as a system parameter passes through a critical point, is
necessarily abrupt. Likewise, the disappearance of multistability is abrupt,
as the typical scenario for a chaotic attractor to be destroyed is through 
a boundary crisis~\cite{GOY:1983}, which is sudden with respect to parameter 
variations. These behaviors are illustrated in Fig.~\ref{fig:SL_p}.

\begin{figure}
\centering
\includegraphics[width=0.8\linewidth]{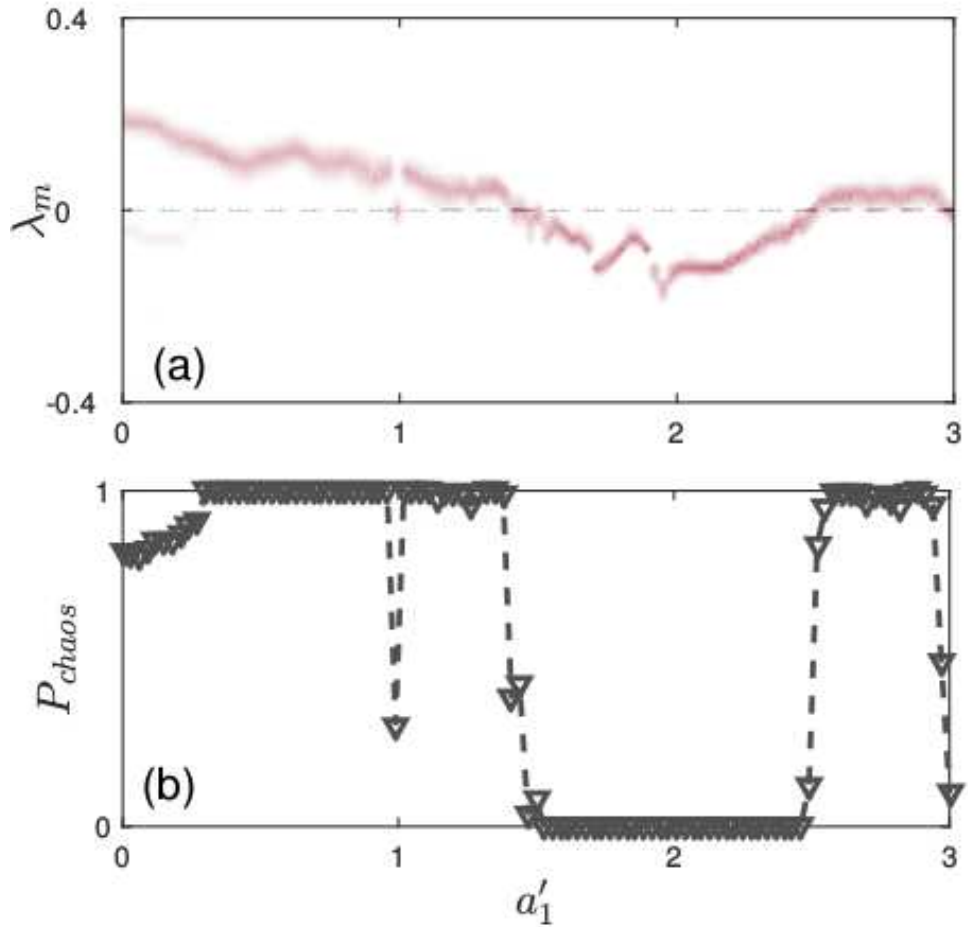}
\caption{{\bf Occurrence of reliable and robust chaos with
probability one under quasiperiodic driving}.
When a second $ac$ driving field of amplitude
$a_1^{\prime}$ and frequency $\Omega^{\prime}=\sqrt{2}$ is applied to the
superlattice system, open parameter intervals emerge in which the probability
of generating chaos from a random initial condition is unity. (a) Statistical
counts of the maximum Lyapunov exponent and (b) probability of generating
chaos versus $a_1^{\prime}$. Other parameters are $a_0=2.23$, $a_1=2.3$,
$a_2 = 0$, $a_3=7.48$, $b_1=0.28$, $b_2=b_3=2.85\times 10^{-2}$, and
$\Omega=1.34$. {\em From Ref.~\cite{YHL:2016} with permission}.}
\label{fig:SL_qp}
\end{figure}

From the point of view of random signal generation, multistability is 
undesired. It was found~\cite{YHL:2016}, however, that an additional
driving field, e.g., of an incommensurate frequency, can effectively eliminate
multistability to guarantee the existence of open parameter regions in which
the probability of generating chaos from random initial conditions is unity.
Such a behavior is illustrated in Fig.~\ref{fig:SL_qp}. It was also 
found~\cite{YHL:2016} that noise plays a similar role in suppressing 
multistability in the sense that weak noise can suppress chaos but strong 
noise can lead to chaos with probability one.

\section{Conclusion}

In nonlinear dynamical systems multistability is ubiquitous, where 
multiple attractors coexist, each with its own basin of attraction.
The basin boundaries separating the distinct basins of attraction 
can be fractal~\cite{GMOY:1983,MGOY:1985} or even riddled~\cite{AYYK:1992,OAKSY:1994,ABS:1994,HCP:1994,LGYV:1996,LG:1996,ABS:1996,LA:2001,Lai:1997,BCP:1997,LG:1999b,Lai:2000}, dynamically leading to transient chaos~\cite{LT:book}.
In applications such as random signal generation in engineering systems 
where robust chaos is required, mitigating or even removing multistability 
so that the system possesses a single chaotic attractor is desired. We have 
laid out in this mini-review the proposition that quasiperiodicity can be 
exploited to suppress and eliminate multistability, and presented an 
experimentally realizable physical system to support the proposition.   

Under what conditions can quasiperiodic forcing suppress and eliminate 
multistability? A heuristic answer is the following. On short time scales,
i.e., in $ t < 1/|\Delta\omega|$, where $|\Delta\omega|$ is the frequency
difference between the two sinusoidal driving signals, the combined effect
of the driving is similar to that due to random noise. If the average 
driving amplitude exceeds the minimum distance from an attractor to its
basin boundary, there is a finite probability that a trajectory can 
exit the basin to approach another coexisting attractor, effectively 
removing the basin of the first attractor. In applications where 
multistability is undesired, e.g., in random signal generation, simply
applying quasiperiodic driving of sufficiently high amplitude can be an 
effective way.

\section*{Acknowledgement}

YCL would like to acknowledge support from the Vannevar Bush
Faculty Fellowship program sponsored by the Basic Research Office of
the Assistant Secretary of Defense for Research and Engineering and
funded by the Office of Naval Research through Grant No.~N00014-16-1-2828.
This work was also supported by ONR through Grant No.~N00014-15-1-2405.


\begin{thebibliography}{100}

\bibitem{PF:1994}
A.~S. Pikovsky, U.~Feudel, Correlations and spectra of strange nonchaotic
  attractors.
\newblock {\it J. Phys. A. Math. Gen.\/} {\bf 27}, 5209-5219 (1994).

\bibitem{FPZ:1995}
U.~Feudel, A.~S. Pikovsky, M.~A. Zaks, Correlation properties of a
  quasiperiodically forced two-level system.
\newblock {\it Phys. Rev. E\/} {\bf 51}, 1762--1769 (1995).

\bibitem{PF:1995}
A.~S. Pikovsky, U.~Feudel, Characterizing strange nonchaotic attractors.
\newblock {\it Chaos\/} {\bf 5}, 253-260 (1995).

\bibitem{KPF:1995}
S.~P. Kuznetsov, A.~S. Pikovsky, U.~Feudel, Birth of a strange nonchaotic
  attractor: A renormalization group analysis.
\newblock {\it Phys. Rev. E\/} {\bf 51}, R1629--R1632 (1995).

\bibitem{FKP:1995}
U.~Feudel, J.~Kurths, A.~S. Pikovsky, Strange non-chaotic attractor in a
  quasiperiodically forced circle map.
\newblock {\it Physica D\/} {\bf 88}, 176-186 (1995).

\bibitem{PZFK:1995}
A.~S. Pikovsky, M.~A. Zaks, U.~Feudel, J.~Kurths, Singular continuous spectra
  in dissipative dynamics.
\newblock {\it Phys. Rev. E\/} {\bf 52}, 285--296 (1995).

\bibitem{LFG:1996}
Y.-C. Lai, U.~Feudel, C.~Grebogi, Scaling behaviors in the transition to chaos
  in quasiperiodically driven dynamical systems.
\newblock {\it Phys. Rev. E\/} {\bf 54}, 6070-6073 (1996).

\bibitem{FPP:1996}
U.~Feudel, A.~Pikovsky, A.~Politi, Renormalization of correlations and spectra
  of a strange non-chaotic attractor.
\newblock {\it J. Phys. A. Math. Gen.\/} {\bf 29}, 5297-5311 (1996).

\bibitem{WFP:1997}
A.~Witt, U.~Feudel, A.~Pikovsky, Birth of strange nonchaotic attractors due to
  interior crisis.
\newblock {\it Physica D\/} {\bf 109}, 180-190 (1997).

\bibitem{FGO:1997}
U.~Feudel, C.~Grebogi, E.~Ott, Phase-locking in quasiperiodically forced
  systems.
\newblock {\it Phys. Rep.\/} {\bf 290}, 11-25 (1997).

\bibitem{FWLG:1998}
U.~Feudel, A.~Witt, Y.-C. Lai, C.~Grebogi, Basin bifurcation in
  quasiperiodically forced systems.
\newblock {\it Phys. Rev. E\/} {\bf 58}, 3060--3066 (1998).

\bibitem{OF:2000}
H.~M. Osinga, U.~Feudel, Boundary crisis in quasiperiodically forced systems.
\newblock {\it Physica D\/} {\bf 141}, 54-64 (2000).

\bibitem{NSMF:2003}
E.~Neumann, I.~Sushko, Y.~Maistrenko, U.~Feudel, Synchronization and
  desynchronization under the influence of quasiperiodic forcing.
\newblock {\it Phys. Rev. E\/} {\bf 67}, 026202 (2003).

\bibitem{SPRF:2005}
M.~D. Shrimali, A.~Prasad, R.~Ramaswamy, U.~Feudel, Basin bifurcations in
  quasiperiodically forced coupled systems.
\newblock {\it Phys. Rev. E\/} {\bf 72}, 036215 (2005).

\bibitem{FKP:book}
U.~Feudel, S.~Kuznetsov, A.~Pikovsky, {\it Strange nonchaotic attractors:
  Dynamics between order and chaos in quasiperiodically forced systems\/}
  (World Scientific, Singapore, 2006).

\bibitem{FGHY:1996}
U.~Feudel, C.~Grebogi, B.~R. Hunt, J.~A. Yorke, Map with more than 100
  coexisting low-period periodic attractors.
\newblock {\it Phys. Rev. E\/} {\bf 54}, 71--81 (1996).

\bibitem{FG:1997}
U.~Feudel, C.~Grebogi, Multistability and the control of complexity.
\newblock {\it Chaos\/} {\bf 7}, 597-604 (1997).

\bibitem{KFG:1999}
S.~Kraut, U.~Feudel, C.~Grebogi, Preference of attractors in noisy multistable
  systems.
\newblock {\it Phys. Rev. E\/} {\bf 59}, 5253--5260 (1999).

\bibitem{KF:2002}
S.~Kraut, U.~Feudel, Multistability, noise, and attractor hopping: The crucial
  role of chaotic saddles.
\newblock {\it Phys. Rev. E\/} {\bf 66}, 015207 (2002).

\bibitem{KF:2003a}
S.~Kraut, U.~Feudel, Enhancement of noise-induced escape through the existence
  of a chaotic saddle.
\newblock {\it Phys. Rev. E\/} {\bf 67}, 015204(R) (2003).

\bibitem{KF:2003b}
S.~Kraut, U.~Feudel, Noise-induced escape through a chaotic saddle: Lowering of
  the activation energy.
\newblock {\it Physica D\/} {\bf 181}, 222-234 (2003).

\bibitem{FG:2003}
U.~Feudel, C.~Grebogi, Why are chaotic attractors rare in multistable systems?
\newblock {\it Phys. Rev. Lett.\/} {\bf 91}, 134102 (2003).

\bibitem{NFS:2011}
C.~N. Ngonghala, U.~Feudel, K.~Showalter, Extreme multistability in a chemical
  model system.
\newblock {\it Phys. Rev. E\/} {\bf 83}, 056206 (2011).

\bibitem{Pateletal:2014}
M.~S. Patel, U.~Patel, S.~Sen, G.~C.~Sethia, C.~Hens, S.~K.~Dana, U.~Feudel, K.~Showalter, C.~N.~Ngonghala, R.~E.~Amritkar,  
Experimental observation of extreme multistability
  in an electronic system of two coupled {R}\"ossler oscillators.
\newblock {\it Phys. Rev. E\/} {\bf 89}, 022918 (2014).

\bibitem{PF:2014}
A.~N. Pisarchik, U.~Feudel, Control of multistability.
\newblock {\it Phys. Rep.\/} {\bf 540}, 167-218 (2014).

\bibitem{DGO:1989}
M.~Ding, C.~Grebogi, E.~Ott, Evolution of attractors in quasiperiodically
  forced systems: From quasiperiodic to strange nonchaotic to chaotic.
\newblock {\it Phys. Rev. A\/} {\bf 39}, 2593--2598 (1989).

\bibitem{GMOY:1983}
C.~Grebogi, S.~W. McDonald, E.~Ott, J.~A. Yorke, Final state sensitivity: an
  obstruction to predictability.
\newblock {\it Phys. Lett. A\/} {\bf 99}, 415-418 (1983).

\bibitem{MGOY:1985}
S.~W. McDonald, C.~Grebogi, E.~Ott, J.~A. Yorke, Fractal basin boundaries.
\newblock {\it Physica D\/} {\bf 17}, 125-153 (1985).

\bibitem{YOC:1990}
L.~Yu, E.~Ott, Q.~Chen, Transition to chaos for random dynamical systems.
\newblock {\it Phys. Rev. Lett.\/} {\bf 65}, 2935-2938 (1990).

\bibitem{YOC:1991}
L.~Yu, E.~Ott, Q.~Chen, Fractal distribution of floaters on a fluid surface and
  the transition to chaos for random maps.
\newblock {\it Physica D\/} {\bf 53}, 102-124 (1991).

\bibitem{RGO:1990}
F.~Romeiras, C.~Grebogi, E.~Ott, Multifractal properties of snapshot attractors
  of random maps.
\newblock {\it Phys. Rev. A\/} {\bf 41}, 784-799 (1990).

\bibitem{LLBS:2002}
Z.~Liu, Y.-C. Lai, L.~Billings, I.~B. Schwartz, Transition to chaos in
  continuous-time random dynamical systems.
\newblock {\it Phys. Rev. Lett.\/} {\bf 88}, 124101 (2002).

\bibitem{LLBS:2003}
Y.-C. Lai, Z.~Liu, L.~Billings, I.~B. Schwartz, Noise-induced unstable
  dimension variability and transition to chaos in random dynamical systems.
\newblock {\it Phys. Rev. E\/} {\bf 67}, 026210 (2003).

\bibitem{TL:2010}
T.~T\'el, Y.-C. Lai, Quasipotential approach to critical scaling in
  noise-induced chaos.
\newblock {\it Phys. Rev. E\/} {\bf 81}, 056208 (2010).

\bibitem{Lai:1996}
Y.-C. Lai, Transition from strange nonchaotic to strange chaotic attractors.
\newblock {\it Phys. Rev. E\/} {\bf 53}, 57--65 (1996).

\bibitem{YL:1996}
T.~Yal\ifmmode\mbox{\c{c}}\else\c{c}\fi{}\ifmmode\imath\else\i\fi{}nkaya, Y.-C.
  Lai, Blowout bifurcation route to strange nonchaotic attractors.
\newblock {\it Phys. Rev. Lett.\/} {\bf 77}, 5039--5042 (1996).

\bibitem{YL:1997}
T.~Yal\ifmmode\mbox{\c{c}}\else\c{c}\fi{}\ifmmode\imath\else\i\fi{}nkaya, Y.-C.
  Lai, Bifurcation to strange nonchaotic attractors.
\newblock {\it Phys. Rev. E\/} {\bf 56}, 1623--1630 (1997).

\bibitem{WZLL:2004}
X.~Wang, M.~Zhan, C.-H. Lai, Y.-C. Lai, Strange nonchaotic attractors in random
  dynamical systems.
\newblock {\it Phys. Rev. Lett.\/} {\bf 92}, 074102 (2004).

\bibitem{Kaneko:1997}
K.~Kaneko, Dominance of {M}ilnor attractors and noise-induced selection in a
  multiattractor system.
\newblock {\it Phys. Rev. Lett.\/} {\bf 78}, 2736-2739 (1997).

\bibitem{YHL:2016}
L.~Ying, D.~Huang, Y.-C. Lai, Multistability, chaos, and random signal
  generation in semiconductor superlattices.
\newblock {\it Phys. Rev. E\/} {\bf 93}, 062204 (2016).

\bibitem{LT:book}
Y.-C. Lai, T.~T\'{e}l, {\it Transient Chaos - Complex Dynamics on Finite Time
  Scales\/} (Springer, New York, 2011).

\bibitem{LGYK:1993}
Y.-C. Lai, C.~Grebogi, J.~A. Yorke, I.~Kan, How often are chaotic saddles
  nonhyperbolic?
\newblock {\it Nonlinearity\/} {\bf 6}, 779-797 (1993).

\bibitem{KGPLS:2002}
H.~Kantz, C.~Grebogi, A.~Prasad, Y.-C. Lai, E.~Sinde, Unexpected robustness
  against noise of a class of nonhyperbolic chaotic attractors.
\newblock {\it Phys. Rev. E\/} {\bf 65}, 026209 (2002).

\bibitem{KG:2004}
S.~Kraut, C.~Grebogi, Escaping from nonhyperbolic chaotic attractors.
\newblock {\it Phys. Rev. Lett.\/} {\bf 92}, 234101 (2004).

\bibitem{SOG:1991}
J.~C. Sommerer, E.~Ott, C.~Grebogi, Scaling law for characteristic times of
  noise-induced crises.
\newblock {\it Phys. Rev. A\/} {\bf 43}, 1754--1769 (1991).

\bibitem{LY:1988}
F.~Ledrappier, L.-S. Young, Dimension formula for random transformations.
\newblock {\it Commun. Math. Phys.\/} {\bf 117}, 529-548 (1988).

\bibitem{FW:book}
M.~I. Freidlin, A.~Wentzell, {\it Random Perturbations of Dynamical Systems\/}
  (Springer, New York, 1984).

\bibitem{GT:1984}
R.~Graham, T.~T\'el, Existence of a potential for dissipative dynamical
  systems.
\newblock {\it Phys. Rev. Lett.\/} {\bf 52}, 9-12 (1984).

\bibitem{Grah:1989}
R.~Graham, {\it Noise in Nonlinear Dynamical Systems, Vol.1\/}, F.~Moss,
  P.~V.~E. McClintock, eds. (Cambridge University Press, Cambridge, 1989), pp.
  225--278.

\bibitem{Gras:1989}
P.~Grassberger, Noise-induced escape from attractors.
\newblock {\it J. Phys. A\/} {\bf 22}, 3283-3290 (1989).

\bibitem{GHT:1991}
R.~Graham, A.~Hamm, T.~T\'el, Nonequilibrium potentials for dynamical systems
  with fractal attractors and repellers.
\newblock {\it Phys. Rev. Lett.\/} {\bf 66}, 3089-3092 (1991).

\bibitem{RT:1991}
P.~Reimann, P.~Talkner, Invariant densities for noisy maps.
\newblock {\it Phys. Rev. A\/} {\bf 44}, 6348-6393 (1991).

\bibitem{HG:1992}
A.~Hamm, R.~Graham, Noise-induced attractor explosions near tangent
  bifurcations.
\newblock {\it J. Stat. Phys.\/} {\bf 66}, 689-725 (1992).

\bibitem{HTG:1994}
A.~Hamm, T.~T\'el, R.~Graham, Noise-induced attractor explosions near tangent
  bifurcations.
\newblock {\it Phys. Lett. A\/} {\bf 185}, 313-320 (1994).

\bibitem{SO:1993}
J.~C. Sommerer, E.~Ott, Particles floating on a moving fluid: a dynamically
  comprehensive physical fractal.
\newblock {\it Science\/} {\bf 259}, 335-339 (1993).

\bibitem{FY:1985}
H.~Fujisaka, T.~Yamada, A new intermittency in coupled dynamical systems.
\newblock {\it Prog. Theo. Phys.\/} {\bf 74}, 918-921 (1985).

\bibitem{FY:1986}
H.~Fujisaka, T.~Yamada, Stability theory of synchronized motion in
  coupled-oscillator systems 4. instability of synchronized chaos and new
  intermittency.
\newblock {\it Prog. Theo. Phys.\/} {\bf 75}, 1087-1104 (1986).

\bibitem{FIIY:1986}
H.~Fujisaka, H.~Ishii, M.~Inoue, T.~Yamada, Intermittency caused by chaotic
  modulation 2. {L}yapunov exponent, fractal structure, and power spectrum.
\newblock {\it Prog. Theo. Phys.\/} {\bf 76}, 1198-1209 (1986).

\bibitem{PST:1993}
N.~Platt, E.~A. Spiegel, C.~Tresser, On-off intermittency: a mechanism for
  bursting.
\newblock {\it Phys. Rev. Lett.\/} {\bf 70}, 279-282 (1993).

\bibitem{HPH:1994}
J.~F. Heagy, N.~Platt, S.~M. Hammel, Characterization of on-off intermittency.
\newblock {\it Phys. Rev. E\/} {\bf 49}, 1140-1150 (1994).

\bibitem{PHH:1994}
N.~Platt, S.~M. Hammel, J.~F. Heagy, Effects of additive noise on on-off
  intermittency.
\newblock {\it Phys. Rev. Lett.\/} {\bf 72}, 3498-3501 (1994).

\bibitem{HPHHL:1994}
P.~W. Hammer, N.~Platt, S.~M. Hammel, J.~F. Heagy, B.~D. Lee, Experimental
  observation of on-off intermittency.
\newblock {\it Phys. Rev. Lett.\/} {\bf 73}, 1095-1098 (1994).

\bibitem{SO:1994}
J.~C. Sommerer, E.~Ott, Blowout bifurcations - the occurrence of riddled basins
  and on-off intermittency.
\newblock {\it Phys. Lett. A\/} {\bf 188}, 39-47 (1994).

\bibitem{LG:1995}
Y.-C. Lai, C.~Grebogi, Intermingled basins and two-state on-off intermittency.
\newblock {\it Phys. Rev. E\/} {\bf 52}, R3313-R3316 (1995).

\bibitem{VAOS:1995}
S.~C. Venkataramani, T.~M. Antonsen, E.~Ott, J.~C. Sommerer, Characterization
  of on-off intermittent time-series.
\newblock {\it Phys. Lett. A\/} {\bf 207}, 173-179 (1995).

\bibitem{AS:1996}
P.~Ashwin, E.~Stone, Influence of noise near blowout bifurcation.
\newblock {\it Phys. Rev. E\/} {\bf 56}, 1635-1641 (1996).

\bibitem{Lai:1996a}
Y.-C. Lai, Symmetry-breaking bifurcation with on-off intermittency in chaotic
  dynamical systems.
\newblock {\it Phys. Rev. E\/} {\bf 53}, R4267-R4270 (1996).

\bibitem{Lai:1996b}
Y.-C. Lai, Distinct small-distance scaling behavior of on-off intermittency in
  chaotic dynamical systems.
\newblock {\it Phys. Rev. E\/} {\bf 54}, 321-327 (1996).

\bibitem{VAOS:1996}
S.~C. Venkataramani, T.~M. Antonsen, E.~Ott, J.~C. Sommerer, On-off
  intermittency - power spectrum and fractal properties of time-series.
\newblock {\it Physica D\/} {\bf 96}, 66-99 (1996).

\bibitem{MALK:2001}
D.~Marthaler, D.~Armbruster, Y.-C. Lai, E.~J. Kostelich, Perturbed on-off
  intermittency.
\newblock {\it Phys. Rev. E\/} {\bf 64}, 016220 (2001).

\bibitem{RC:2007}
E.~L. Rempel, A.~C. Chian, Origin of transient and intermittent dynamics in
  spatiotemporal chaotic systems.
\newblock {\it Phys. Rev. Lett.\/} {\bf 98}, 014101 (2007).

\bibitem{HNDSCL:2017}
L.~Huang, X.~Ni, W.~L.~Ditto, M.~Spano, P.~R.~Carney, Y.-C.~Lai,
Detecting and characterizing high frequency
  oscillations in epilepsy - a case study of big data analysis.
\newblock {\it Roy. Soc. Open Sci.\/} {\bf 4}, 160741 (2017).

\bibitem{SLC:2008}
R.~Serquina, Y.-C. Lai, Q.~Chen, Characterization of nonstationary chaotic
  systems.
\newblock {\it Phys. Rev. E\/} {\bf 77}, 026208 (2008).

\bibitem{KFT:2016}
B.~Kasz\'as, U.~Feudel, T.~T\'el, Death and revival of chaos.
\newblock {\it Phys. Rev. E\/} {\bf 94}, 062221 (2016).

\bibitem{RBOAG:1987}
F.~J. Romeiras, A.~Bondeson, E.~Ott, T.~M.~A. Jr, C.~Grebogi, Quasiperiodically
  forced dynamical systems with strange nonchaotic attractors.
\newblock {\it Physica D\/} {\bf 26}, 277-294 (1987).

\bibitem{RO:1987}
F.~J. Romeiras, E.~Ott, Strange nonchaotic attractors of the damped pendulum
  with quasiperiodic forcing.
\newblock {\it Phys. Rev. A\/} {\bf 35}, 4404--4413 (1987).

\bibitem{BOA:1985}
A.~Bondeson, E.~Ott, T.~M. Antonsen, Quasiperiodically forced damped pendula
  and {S}chr\"odinger equations with quasiperiodic potentials: Implications of
  their equivalence.
\newblock {\it Phys. Rev. Lett.\/} {\bf 55}, 2103--2106 (1985).

\bibitem{ET:1970}
L.~Esaki, R.~Tsu, Superlattice and negative differential conductivity in
  semiconductors.
\newblock {\it IBM J. Res. Dev.\/} {\bf 14}, 61-65 (1970).

\bibitem{Grahn:book}
H.~T. Grahn, {\it Semiconductor Supperlattices, Growth and Electronic
  Properties\/} (World Scientific, Singapore, 1995).

\bibitem{ZKKPG:1996}
Y.~Zhang, J.~Kastrup, R.~Klann, K.~H. Ploog, H.~T. Grahn, Synchronization and
  chaos induced by resonant tunneling in gaas/alas superlattices.
\newblock {\it Phys. Rev. Lett.\/} {\bf 77}, 3001--3004 (1996).

\bibitem{LT:1984}
X.~L. Lei, C.~S. Ting, Theory of nonlinear electron transport for solids in a
  strong electric field.
\newblock {\it Phys. Rev. B\/} {\bf 30}, 4809--4812 (1984).

\bibitem{LT:1985}
X.~L. Lei, C.~S. Ting, Green's-function approach to nonlinear electronic
  transport for an electron-impurity-phonon system in a strong electric field.
\newblock {\it Phys. Rev. B\/} {\bf 32}, 1112--1132 (1985).

\bibitem{LHC:1991}
X.~L. Lei, N.~J.~M. Horing, H.~L. Cui, Theory of negative differential
  conductivity in a superlattice miniband.
\newblock {\it Phys. Rev. Lett.\/} {\bf 66}, 3277--3280 (1991).

\bibitem{Lei:1994a}
X.~L. Lei, High-frequency differential mobility in vertical transport of a
  confined superlattice.
\newblock {\it J. Phys. Cond. Matt.\/} {\bf 6}, 10043 (1994).

\bibitem{Lei:1994b}
X.~L. Lei, Balance equations for electron transport in a general energy band.
\newblock {\it J. Phys. Cond. Matt.\/} {\bf 6}, 9189 (1994).

\bibitem{IDS:1991}
A.~A. Ignatov, E.~P. Dodin, V.~I. Shashkin, Transient response theory for
  semiconductor superlattices: connection with {B}loch oscillations.
\newblock {\it Mod. Phys. Lett. B\/} {\bf 5}, 1087 (1991).

\bibitem{Gerhardts:1993}
R.~R. Gerhardts, Effect of elastic scattering on miniband transport in
  semiconductor superlattices.
\newblock {\it Phys. Rev. B\/} {\bf 48}, 9178--9181 (1993).

\bibitem{BT:1977}
M.~B\"uttiker, H.~Thomas, Current instability and domain propagation due to
  {B}ragg scattering.
\newblock {\it Phys. Rev. Lett.\/} {\bf 38}, 78--80 (1977).

\bibitem{BT1:1979}
M.~B\"uttiker, H.~Thomas, Hydrodynamic modes, soft modes and fluctuation
  spectra near the threshold of a current instability.
\newblock {\it Z. Phys. B\/} {\bf 33}, 275--287 (1979).

\bibitem{BT2:1979}
M.~B\"uttiker, H.~Thomas, Bifurcation and stability of dynamical structures at
  a current instability.
\newblock {\it Z. Phys. B\/} {\bf 34}, 301--311 (1979).

\bibitem{Lei:1992}
X.~L. Lei, Balance equations for hot electron transport in an arbitrary energy
  band.
\newblock {\it Phys. Status Solidi B\/} {\bf 170}, 519-529 (1992).

\bibitem{Lei:1995}
X.~L. Lei, Investigation of the {B}uttiker-{T}homas momentum balance equation
  from the {H}eisenberg equation of motion for {B}loch electrons.
\newblock {\it J. Phys. Conden. Matt.\/} {\bf 7}, L429 (1995).

\bibitem{ACMKC:1998}
K.~N. Alekseev, E.~H. Cannon, J.~C. McKinney, F.~V. Kusmartsev, D.~K. Campbell,
  Spontaneous dc current generation in a resistively shunted semiconductor
  superlattice driven by a terahertz field.
\newblock {\it Phys. Rev. Lett.\/} {\bf 80}, 2669--2672 (1998).

\bibitem{ABCCC:1996}
K.~N. Alekseev, G.~P. Berman, D.~K. Campbell, E.~H. Cannon, M.~C. Cargo,
  Dissipative chaos in semiconductor superlattices.
\newblock {\it Phys. Rev. B\/} {\bf 54}, 10625--10636 (1996).

\bibitem{Scholl:book}
E.~Sch\"{o}ll, {\it Nonlinear Spatiotemporal Dynamics and Chaos in
  Semiconductors\/} (Cambridge University Press, Cambridge, UK, 2001).

\bibitem{BGB:1996}
O.~M. Bulashenko, M.~J. Garc\'{\i}a, L.~L. Bonilla, Chaotic dynamics of
  electric-field domains in periodically driven superlattices.
\newblock {\it Phys. Rev. B\/} {\bf 53}, 10008--10018 (1996).

\bibitem{PSS:1998}
M.~Patra, G.~Schwarz, E.~Sch\"oll, Bifurcation analysis of stationary and
  oscillating domains in semiconductor superlattices with doping fluctuations.
\newblock {\it Phys. Rev. B\/} {\bf 57}, 1824--1833 (1998).

\bibitem{LGPB:1998}
K.~J. Luo, H.~T. Grahn, K.~H. Ploog, L.~L. Bonilla, Explosive bifurcation to
  chaos in weakly coupled semiconductor superlattices.
\newblock {\it Phys. Rev. Lett.\/} {\bf 81}, 1290--1293 (1998).

\bibitem{ACMKC2:1998}
K.~N. Alekseev, E.~H. Cannon, J.~C. McKinney, F.~V. Kusmartsev, D.~K. Campbell,
  Symmetry-breaking and chaos in electron transport in semiconductor
  superlattices.
\newblock {\it Physica D\/} {\bf 113}, 129--133 (1998).

\bibitem{CL:1999}
J.~C. Cao, X.~L. Lei, Synchronization and chaos in miniband semiconductor
  superlattices.
\newblock {\it Phys. Rev. B\/} {\bf 60}, 1871--1878 (1999).

\bibitem{CLL:2000}
J.~C. Cao, H.~C. Liu, X.~L. Lei, Chaotic dynamics in quantum-dot miniband
  superlattices.
\newblock {\it Phys. Rev. B\/} {\bf 61}, 5546--5555 (2000).

\bibitem{SPB:2001}
D.~S\'anchez, G.~Platero, L.~L. Bonilla, Quasiperiodic current and strange
  attractors in ac-driven superlattices.
\newblock {\it Phys. Rev. B\/} {\bf 63}, 201306 (2001).

\bibitem{AK:2002}
K.~N. Alekseev, F.~V. Kusmartsev, Pendulum limit, chaos and phase-locking in
  the dynamics of ac-driven semiconductor superlattices.
\newblock {\it Phys. Lett. A\/} {\bf 305}, 281--288 (2002).

\bibitem{ABC:2002}
K.~N. Alekseev, G.~P. Bermana, D.~K. Campbell, Dynamical instabilities and
  deterministic chaos in ballistic electron motion in semiconductor
  superlattices.
\newblock {\it Phys. Lett. A\/} {\bf 193}, 54-60 (1994).

\bibitem{ASWS:2002}
A.~Amann, J.~Schlesner, A.~Wacker, E.~Sch\"oll, Chaotic front dynamics in
  semiconductor superlattices.
\newblock {\it Phys. Rev. B\/} {\bf 65}, 193313 (2002).

\bibitem{BG:2005}
L.~L. Bonilla, H.~T. Grahn, Non-linear dynamics of semiconductor superlattices.
\newblock {\it Rep. Prog. Phys.\/} {\bf 68}, 577-683 (2005).

\bibitem{HS:2008}
J.~Gal\'{a}n, L.~L. Bonilla, M.~Moscoso, Bifurcation behavior of a superlattice
  model.
\newblock {\it SIAM J. Appl. Math.\/} {\bf 60}, 2029-2057 (2006).

\bibitem{GBSF:2009}
M.~T. Greenaway, A.~G. Balanov, E.~Sch\"oll, T.~M. Fromhold, Controlling and
  enhancing terahertz collective electron dynamics in superlattices by
  chaos-assisted miniband transport.
\newblock {\it Phys. Rev. B\/} {\bf 80}, 205318 (2009).

\bibitem{Stapletonetal:2004}
S.~P. Stapleton, S.~Bujkiewicz, T.~M.~Fromhold, P.~B.~Wilkinson, A.~Patan\'{e}, L.~Eaves, A.~A.~Krokhin, M.~Henini, N.~S.~Sankeshwar, F.~W.~Sheard, 
Use of stochastic web patterns to control
electron transport in semiconductor superlattices.
\newblock {\it Physica D\/} {\bf 199}, 166-172 (2004).

\bibitem{WC:2005}
C.~Wang, J.-C. Cao, Current oscillation and chaotic dynamics in superlattices
  driven by crossed electric and magnetic fields.
\newblock {\it Chaos\/} {\bf 15}, 013111 (2005).

\bibitem{BFPEF:2008}
A.~G. Balanov, D.~Fowler, A.~Patan\`e, L.~Eaves, T.~M. Fromhold, Bifurcations
  and chaos in semiconductor superlattices with a tilted magnetic field.
\newblock {\it Phys. Rev. E\/} {\bf 77}, 026209 (2008).

\bibitem{LGTP:1998}
K.~J. Luo, H.~T. Grahn, S.~W. Teitsworth, K.~H. Ploog, Influence of higher
  harmonics on poincar\'e maps derived from current self-oscillations in a
  semiconductor superlattice.
\newblock {\it Phys. Rev. B\/} {\bf 58}, 12613--12616 (1998).

\bibitem{ZKGP:1997}
Y.-H. Zhang, R.~Klann, H.~T. Grahn, K.~H. Ploog, Transition between
  synchronization and chaos in doped gaas/alas superlattices.
\newblock {\it Superlatt. Microstruc.\/} {\bf 21}, 565--568 (1997).

\bibitem{Fromholdetal:2004}
T.~M. Fromhold, A.~Patan\'{e}, S.~Bujkiewicz, P.~B.~Wilkinson, D.~Fowler, D.~Sherwood, S.~P.~Stapleton, A.~A.~Krokhin, L.~Eaves, M.~Henini, N.~S.~Sankeshwar, F.~W.~Sheard, Chaotic electron diffusion through stochastic
  webs enhances current flow in superlattices.
\newblock {\it Nature\/} {\bf 428}, 726--730 (2004).

\bibitem{Hramovetal:2014}
A.~E. Hramov, V.~V.~Makarov, A.~A.~Koronovskii, S.~A.~Kurkin, M.~B.~Gaifullin, N.~V.~Alexeeva, K.~N.~Alekseev, M.~T.~Greenaway, T.~M.~Fromhold, A.~Patan\`e, F.~.V.~Kusmartsev, V.~A.~Maksimenko, O.~I.~Moskalenko, A.~G.~Balanov, 
Subterahertz chaos generation by coupling a
  superlattice to a linear resonator.
\newblock {\it Phys. Rev. Lett.\/} {\bf 112}, 116603 (2014).

\bibitem{HSA:2007}
T.~Hyart, A.~V. Shorokhov, K.~N. Alekseev, Theory of parametric amplification
  in superlattices.
\newblock {\it Phys. Rev. Lett.\/} {\bf 98}, 220404 (2007).

\bibitem{HAT:2008}
T.~Hyart, K.~N. Alekseev, E.~V. Thuneberg, {B}loch gain in dc-ac-driven
  semiconductor superlattices in the absence of electric domains.
\newblock {\it Phys. Rev. B\/} {\bf 77}, 165330 (2008).

\bibitem{HAMA:2009}
T.~Hyart, N.~V. Alexeeva, J.~Mattas, K.~N. Alekseev, Terahertz {B}loch
  oscillator with a modulated bias.
\newblock {\it Phys. Rev. Lett.\/} {\bf 102}, 140405 (2009).

\bibitem{HMA:2009}
T.~Hyart, J.~Mattas, K.~N. Alekseev, Model of the influence of an external
  magnetic field on the gain of terahertz radiation from semiconductor
  superlattices.
\newblock {\it Phys. Rev. Lett.\/} {\bf 103}, 117401 (2009).

\bibitem{Hyart:book}
T.~Hyart, {\it Tunable Superlattice Amplifiers Based on Dynamics of Miniband
  Electrons in Electric and Magnetic Fields (Ph.D. Dissertation)\/} (University
  of Oulu, Finland, 2009).

\bibitem{Siegel:2002}
P.~H. Siegel, Terahertz technology.
\newblock {\it IEEE Trans. Microwave Theory Tech.\/} {\bf 50}, 910-928 (2002).

\bibitem{FZ:2002}
B.~Ferguson, X.-C. Zhang, Materials for terahertz science and technology.
\newblock {\it Nat. Mater.\/} {\bf 1}, 26-33 (2002).

\bibitem{CBPHW:2005}
T.~W. Crowe, W.~L. Bishop, D.~W. Porterfield, J.~L. Hesler, R.~M. Weikle,
  Opening the terahertz window with integrated diode circuits.
\newblock {\it IEEE J. Solid-State Cir.\/} {\bf 40}, 2104-2110 (2005).

\bibitem{Tonouchi:2009}
M.~Tonouchi, Cutting-edge terahertz technology.
\newblock {\it Nat. Photon.\/} {\bf 1}, 97-105 (2009).

\bibitem{Kocarev:2001}
L.~Kocarev, Chaos-based cryptography: a brief overview.
\newblock {\it IEEE Cir. Sys. Magaz.\/} {\bf 1}, 6-21 (2001).

\bibitem{SK:2001}
T.~Stojanovski, L.~Kocarev, Chaos-based random number generators-part i:
  analysis [cryptography].
\newblock {\it IEEE Trans. Cir. Sys. I. Funda. Theo. App.\/} {\bf 48}, 281-288
  (2001).

\bibitem{SPK:2001}
T.~Stojanovski, , J.~Pihl, L.~Kocarev, Chaos-based random number generators.
  part ii: practical realization.
\newblock {\it IEEE Trans. Cir. Sys. I. Funda. Theo. App.\/} {\bf 48}, 382-385
  (2001).

\bibitem{DG:2006}
M.~Drutarovsk\'{y}, P.~Galajda, Chaos-based true random number generator
  embedded in a mixed-signal reconfigurable hardware.
\newblock {\it J. Elec. Eng.\/} {\bf 57}, 218-225 (2006).

\bibitem{LC:2006}
T.~Lin, L.~O. Chua, A new class of pseudo-random number generator based on
  chaos in digital filters.
\newblock {\it Int. J. Cir. Theo. App.\/} {\bf 21}, 473-480 (2006).

\bibitem{Uchidaetal:2008}
A.~Uchida, K.~Amano, M.~Inoue, K.~Hirano, S.~Naito, H.~Someya, I.~Oowada, T.~Kurashige, M.~Shiki, S.~Yoshimori, K.~Yoshimura, P.~Davis,
Fast physical random bit generation with chaotic
  semiconductor lasers.
\newblock {\it Nat. Photon.\/} {\bf 2}, 728-732 (2008).

\bibitem{RARK:2009}
I.~Reidler, Y.~Aviad, M.~Rosenbluh, I.~Kanter, Ultrahigh-speed random number
  generation based on a chaotic semiconductor laser.
\newblock {\it Phys. Rev. Lett.\/} {\bf 103}, 024102 (2009).

\bibitem{CHLGD:2010}
Q.~Chen, L.~Huang, Y.-C. Lai, C.~Grebogi, D.~Dietz, Extensively chaotic motion
  in electrostatically driven nanowires and applications.
\newblock {\it Nano lett.\/} {\bf 10}, 406--413 (2010).

\bibitem{ZWLXLWZ:2012}
J.-Z. Zhang, Y.-C.~Wang, M.~Liu, L.-G.~Xue, P.~Li, A.-B.~Wang, M.-J.~Zhang, 
A robust random number generator based on
  differential comparison of chaotic laser signals.
\newblock {\it Opt. Expr.\/} {\bf 20}, 7496-7506 (2012).

\bibitem{HAAC:2005}
D.~Huang, P.~M. Alsing, T.~Apostolova, D.~A. Cardimona, Coupled energy-drift
  and force-balance equations for high-field hot-carrier transport.
\newblock {\it Phys. Rev. B\/} {\bf 71}, 195205 (2005).

\bibitem{HA:2008}
D.~Huang, P.~M. Alsing, Many-body effects on optical carrier cooling in
  intrinsic semiconductors at low lattice temperatures.
\newblock {\it Phys. Rev. B\/} {\bf 78}, 035206 (2008).

\bibitem{HC:2003}
D.~Huang, D.~A. Cardimona, Nonadiabatic effects in a self-consistent hartree
  model for electrons under an ac electric field in multiple quantum wells.
\newblock {\it Phys. Rev. B\/} {\bf 67}, 245306 (2003).

\bibitem{HLG:2009}
D.~Huang, S.~K. Lyo, G.~Gumbs, {B}loch oscillation, dynamical localization, and
  optical probing of electron gases in quantum-dot superlattices in high
  electric fields.
\newblock {\it Phys. Rev. B\/} {\bf 79}, 155308 (2009).

\bibitem{GOY:1983}
C.~Grebogi, E.~Ott, J.~A. Yorke, Crises, sudden changes in chaotic attractors
  and transient chaos.
\newblock {\it Physica D\/} {\bf 7}, 181-200 (1983).

\bibitem{AYYK:1992}
J.~C. Alexander, J.~A. Yorke, Z.~You, I.~Kan, Riddled basins.
\newblock {\it Int. J. Bifur. Chaos Appl. Sci. Eng.\/} {\bf 2}, 795-813 (1992).

\bibitem{OAKSY:1994}
E.~Ott, J.~C. Alexander, I.~Kan, J.~C. Sommerer, J.~A. Yorke, The transition to
  chaotic attractors with riddled basins.
\newblock {\it Physica D\/} {\bf 76}, 384-410 (1994).

\bibitem{ABS:1994}
P.~Ashwin, J.~Buescu, I.~Stewart, Bubbling of attractors and synchronisation of
  oscillators.
\newblock {\it Phys. Lett. A\/} {\bf 193}, 126-139 (1994).

\bibitem{HCP:1994}
J.~F. Heagy, T.~L. Carroll, L.~M. Pecora, Experimental and numerical evidence
  for riddled basins in coupled chaotic systems.
\newblock {\it Phys. Rev. Lett.\/} {\bf 73}, 3528-3531 (1994).

\bibitem{LGYV:1996}
Y.-C. Lai, C.~Grebogi, J.~A. Yorke, S.~Venkataramani, Riddling bifurcation in
  chaotic dynamical systems.
\newblock {\it Phys. Rev. Lett.\/} {\bf 77}, 55-58 (1996).

\bibitem{LG:1996}
Y.-C. Lai, C.~Grebogi, Noise-induced riddling in chaotic dynamical systems.
\newblock {\it Phys. Rev. Lett.\/} {\bf 77}, 5047-5050 (1996).

\bibitem{ABS:1996}
P.~Ashwin, J.~Buescu, I.~Stewart, From attractor to chaotic saddle: a tale of
  transverse instability.
\newblock {\it Nonlinearity\/} {\bf 9}, 703-737 (1996).

\bibitem{LA:2001}
Y.-C. Lai, V.~Andrade, Catastrophic bifurcation from riddled to fractal basins.
\newblock {\it Phys. Rev. E\/} {\bf 64}, 056228 (2001).

\bibitem{Lai:1997}
Y.-C. Lai, Scaling laws for noise-induced temporal riddling in chaotic systems.
\newblock {\it Phys. Rev. E\/} {\bf 56}, 3897-3908 (1997).

\bibitem{BCP:1997}
L.~Billings, J.~H. Curry, E.~Phipps, Lyapunov exponents, singularities, and a
  riddling bifurcation.
\newblock {\it Phys. Rev. Lett.\/} {\bf 79}, 1018-1021 (1997).

\bibitem{LG:1999b}
Y.-C. Lai, C.~Grebogi, Riddling of chaotic sets in periodic windows.
\newblock {\it Phys. Rev. Lett.\/} {\bf 83}, 2926-2929 (1999).

\bibitem{Lai:2000}
Y.-C. Lai, Catastrophe of riddling.
\newblock {\it Phys. Rev. E\/} {\bf 62}, R4505--R4508 (2000).

\end{thebibliography}

\end{document}